\documentclass{SciPost}

\usepackage{amssymb}
\usepackage{xcolor}

\begin{document}

\begin{center}{\Large \textbf{Observation of a power transfer controlled by the phase of a quantum superposition
}}\end{center}

\begin{center}
N. Cottet\textsuperscript{1,2$\dagger$},
S. Jezouin\textsuperscript{1,2$\dagger$},
B. Reznychenko\textsuperscript{3$\ddagger$},
A. Auff\`eves\textsuperscript{3,4,5},
B. Huard\textsuperscript{2,1,6$\star$}
\end{center}

\begin{center}
{\bf 1} Laboratoire de Physique de l'\'Ecole normale sup\'erieure, ENS, Universit\'e PSL,
CNRS, Sorbonne Universit\'e, Universit\'e Paris Cit\'e, F-75005 Paris, France
\\
{\bf 2} \'Ecole Normale Sup\'erieure de Lyon,  CNRS, Laboratoire de Physique, F-69342 Lyon, France
\\
{\bf 3} CNRS and Universit\'e Grenoble Alpes, Institut N\'eel, F-38042 Grenoble, France
\\
{\bf 4} MajuLab, CNRS-UCA-SU-NUS-NTU International Joint Research Laboratory
\\
{\bf 5} Center for Quantum Technologies, National University of Singapore, 117543 Singapore, Singapore
\\
{\bf 6} International Research Laboratory, French American Center for Theoretical Science, CNRS, KITP, Santa Barbara, 93106-4030 USA.
\\
[\baselineskip]
$\dagger$ {\small Present address: Alice \& Bob, 53 Bd du G\'en\'eral Martial Valin, 75015 Paris, France}\\
$\ddagger$ {\small Present address: Photonic Inc., Coquitlam, British Columbia, Canada}\\ $\star$ \href{mailto:benjamin.huard@ens-lyon.fr}{\small benjamin.huard@ens-lyon.fr}
\end{center}

\begin{center}
\today
\end{center}

\section*{Abstract}
{\bf
A driven qubit exchanges energy with the propagating modes that drive it. When two spatially separated modes drive a single qubit with opposite amplitudes, their net action on the qubit cancels. Yet the qubit can still transfer power from one mode to the other through stimulated emission. The directionality originates from opposite stimulated emission powers into each line. We realize this situation with a superconducting transmon qubit coupled to two transmission lines and show that the direction of the power flow is set by the phase $\phi$ of the qubit superposition between its ground and excited states, rather than by any classical control parameter. From a time-resolved measurement of the output power in one line, we observe a transfer that varies as $\cos\phi$ and hence changes direction between $\phi=0$ and $\phi=\pi$. The directionality of the total power flow is limited by the phase independent contributions of the reflected drive and of spontaneous emission, which sets a routing efficiency that we measure as a function of the input power. For an equal superposition of ground and excited qubit states (maximal coherence), the efficiency reaches $63\%$, close to the bound of $69\%$ expected from the measured qubit coherence times.
}


\section{Introduction}
\label{sec:intro}

When considering an open quantum system, the drives are usually modeled as classical parameters in the dynamics. Yet, when the average internal energy of the quantum system varies under the action of a drive, conservation of energy dictates that the energy of the driving mode must change as well. Quantum thermodynamics and quantum energetics focus on these flows of energy and information between the system and its environment~\cite{Binder2018a,AuffevesElouard2026QuantumEnergetics,Campbell2026}, and in particular on the energy flows from one mode of the environment to another and on how they relate to the inner state of the quantum system. It has for instance been shown that the heat flowing across a Josephson junction depends on its superconducting phase difference~\cite{Timossi2018} and that a superconducting qubit can gate the photonic heat current between two reservoirs~\cite{Ronzani2018}. In parallel, superconducting artificial atoms coupled to a one-dimensional waveguide can route microwave photons between propagating modes~\cite{Hoi2011}, a capability that belongs to the broader framework of chiral quantum optics~\cite{Lodahl2017}. At optical frequencies, single-emitter realizations of photon routing have been demonstrated with one atom coupled to a microresonator that sorts and redirects single photons~\cite{Aoki2009}, with an all-optical switch whose output port is set by the state of the atom and toggled by a single photon~\cite{Shomroni2014}, and with the internal state of an emitter controlling the propagation direction of the light it radiates or scatters, yielding nanophotonic isolators and circulators operated by a single atom~\cite{Mitsch2014,Sayrin2015,Scheucher2016}. Several schemes have been proposed to make a superconducting emitter directional in a waveguide~\cite{Gheeraert2020,Guimond2020}, and on-demand directional scattering has been demonstrated with a pair of modulated transmons whose relative modulation phase sets the scattering direction~\cite{Redchenko2023}.

In these experiments the energy and photon flow is governed by a classical control parameter, such as the magnetic flux threading a superconducting loop in the thermal router~\cite{Timossi2018}, the engineered phases of the control fields used to set the direction of single-photon emission~\cite{Kannan2023}, or the adiabatic modulation parameters that fix a topologically quantized power transfer between modes of a superconducting circuit~\cite{Luneau2022}. Even in the single-emitter routers mentioned above, where the emitter itself selects the output, the control variable is its internal state prepared as a classical which-state bit rather than the phase of a coherent superposition. Here we ask instead whether the direction of the energy flow between two modes can be set by a genuinely quantum parameter of the system, namely the phase of a coherent superposition of qubit states.

To this end, we design and realize an experiment where a single qubit acts as an energy pump between two spatially separated propagating electromagnetic modes, with the direction of the power flow controlled by the phase of the superposition between its ground $|g\rangle$ and excited $|e\rangle$ states. By directly measuring the time evolution of the output power on one of the qubit ports we demonstrate that the quantum coherence of the qubit acts as a non-classical knob for power routing, which we quantify through a photon routing efficiency whose evolution we study with the amplitude of the incoming electromagnetic drives. We measure a photon routing efficiency of 63\% for a qubit initialized in an equal superposition of $|e\rangle$ and $|g\rangle$, close to the theoretical limit of 69\%. This limit is computed using a measurement of the qubit coupling to the propagating mode (Purcell rate) using fluorescence. Our results add to the growing ensemble of thermodynamic machines operating in the quantum regime~\cite{Aamir2025} and, more broadly, to the study of energy transfers between light and matter in the elementary setting of a two-level system coupled to a waveguide~\cite{Maffei2021,Wenniger2023,Prasad2024}. In this setting a qubit coupled to a waveguide can act as a charger for the propagating field seen as a battery, the work being the coherent part of the radiated field~\cite{Monsel2020}, or as a measurement-powered engine whose delivered power is set by the steady-state coherence of the qubit~\cite{ElouardJordan2018,Elouard2017PRL,Dassonneville2026}.

The article is organized as follows. Section~\ref{sec:direct} introduces the principle of the experiment and derives the power emitted by the qubit in each propagating mode. Section~\ref{sec:purcell} is devoted to the experimental measurement of the Purcell rate between the qubit and the transmission line in which fluorescence is detected. We demonstrate photon routing in Section~\ref{sec:routing} and show how it is controlled by the phase of the quantum superposition of the qubit states. Finally, we define and measure the transfer efficiency in Section~\ref{sec:efficiency}.

\section{\label{sec:direct}Directionality using stimulated emission}
\begin{figure}[h!]
\centering
\includegraphics[width=8.5cm]{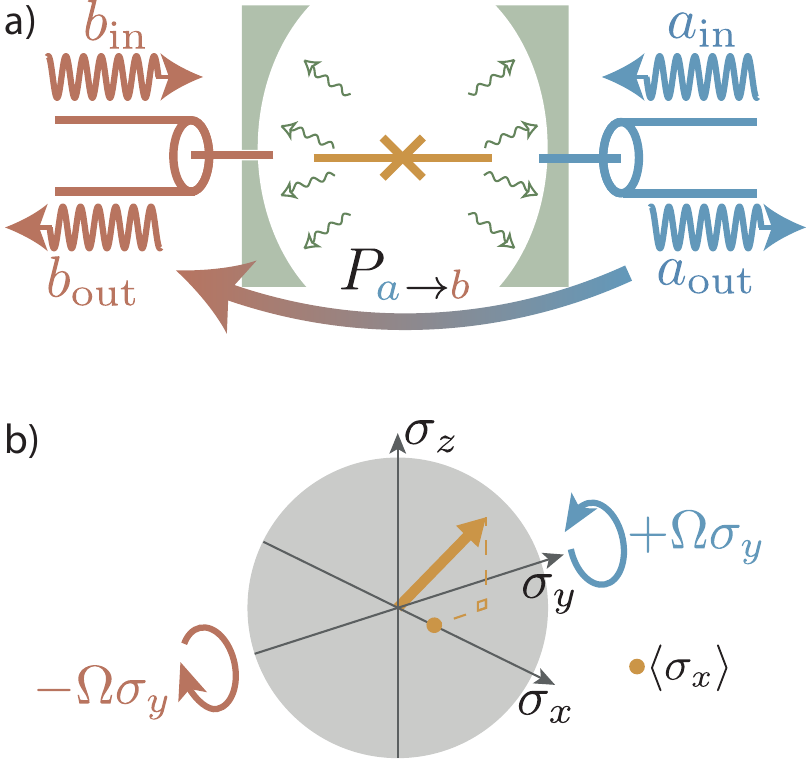}
\caption{\textbf{Power transfer mediated by the phase of a qubit.} (a) A transmon superconducting qubit is resonantly driven through two transmission lines $a$ and $b$ leading to a power transfer $P_{a\rightarrow b}$ between them. (b) When the driving fields are set so that they drive the qubit with the same amplitude but opposite phases (here around $\sigma_y)$, their net effect on the qubit vanishes. Still, the transferred power from one line into the other depends on the qubit coherence, here it is proportional to $\langle\sigma_x\rangle$.}
\label{Fig1}
\end{figure}

Stimulated emission of a qubit into the electromagnetic mode that drives it contributes to the work transfer between the qubit and that mode~\cite{Cottet2017,Monsel2020,Prasad2024,Potts2025}; it can also amplify a probe field, as demonstrated with a single artificial atom at the end of a waveguide~\cite{Aziz2025}. More precisely, stimulated emission is positive when the qubit is driven from a higher energy state to a lower energy state and corresponds to work being transferred from the qubit to the mode. Conversely, driving the qubit from a low-energy state to a high-energy one results in negative stimulated emission, which is absorption of work from the drive. More general operational definitions of work in this waveguide setting can also attribute a contribution to spontaneous emission~\cite{Prasad2024,Potts2025}; here the routed power is carried by the coherent, stimulated contribution. Let us now consider an experiment where the qubit is driven by two spatially separated modes at the same time. The complex amplitudes of the drives can be set such that the apparent field seen by the qubit vanishes, so that in the absence of decoherence the qubit state would remain constant. From the point of view of the modes, this occurs when one drive extracts exactly the same amount of instantaneous work power from the qubit that the other provides (see Fig.~\ref{Fig1}). Thus, counter-intuitively, the qubit can act as a pump of power from one mode to the other, as we compute below, while the net effect of the simultaneous driving on its dynamics is null. Seeing the qubit as inert is thus an illusion: it continuously transfers power from one field to the other, reminiscent of the Zeno regime of Ref.~\cite{Elouard2017PRL} where a qubit kept frozen by measurement still converts energy into a coherent field radiated in the waveguide.

Our experimental implementation consists of a transmon superconducting qubit with transition frequency $\omega_q/2\pi=7.088~\mathrm{GHz}$~\cite{Paik2011a,Cottet2017} coupled to two transmission lines $a$ and $b$ via a far-detuned microwave cavity with the respective coupling rates $\gamma_a$ and $\gamma_b$ (see Fig.~\ref{Fig1}a). The qubit is coupled to the microwave modes at frequency $\omega_q$ that propagate in the two transmission lines. For each line, we distinguish incoming and outgoing modes, so that we denote the four modes of interest by their annihilation operators $a_\mathrm{in}(t)$, $a_\mathrm{out}(t)$, $b_\mathrm{in}(t)$ and $b_\mathrm{out}(t)$ in a scattering picture~\cite{Gardiner2004,Roy2018}. The cavity plays no role in the dynamics of the modes we focus on, apart from determining the Purcell coupling rates between the qubit and the transmission lines $\gamma_{a,b}$ and is only used to characterize the qubit properties. We characterize the output field $a_\mathrm{out}$ by a continuous heterodyne measurement after quantum-limited amplification~\cite{PhysRevX.6.011002}. In order to collect as much fluorescence signal as possible in line $a$ we design two asymmetric coupling rates with $\gamma_b\approx \gamma_a/100$.

Treating the qubit as a lossless emitter coupled only to the two lines (the effect of intrinsic losses is addressed in Sec.~\ref{sec:purcell}), the two outgoing modes $a_\mathrm{out}$ and $b_\mathrm{out}$ can be expressed using the input-output relations for each transmission line~\cite{Gardiner2004} and read
\begin{equation}
a_\mathrm{out} = a_\mathrm{in} - \sqrt{\gamma_a} \sigma_-  \text{ and } 
b_\mathrm{out}= b_\mathrm{in} - \sqrt{\gamma_b} \sigma_-
\label{equ:input}
\end{equation}
with $\sigma_-=|g\rangle\langle e|$ the annihilation operator of the qubit. The choice of phase between the input drives $a_\mathrm{in},b_\mathrm{in}$ reflected on the far-detuned cavity and the fluorescence fields emitted by the qubit $\sqrt{\gamma_{a,b}}\sigma_-$ sets the relative phase between the qubit reference frame and the drives. In the following we only consider the situation where the drives are either in phase or in opposition of phase from each other. Thus, without loss of generality, one can assert $\langle a_\mathrm{in}\rangle, \langle b_\mathrm{in}\rangle\in\mathbb{R}$ and write the qubit Hamiltonian in the rotating frame and after Rotating Wave Approximation as
\begin{equation}
    H = -\hbar(\Omega_a+\Omega_b)\sigma_y/2\ .
\end{equation} 
Here $\Omega_{a}=2\sqrt{\gamma_a}\langle a_\mathrm{in}\rangle$ (resp. $\Omega_{b}=2\sqrt{\gamma_b}\langle b_\mathrm{in}\rangle$) is the Rabi frequency induced by the drive in $a$ (resp. $b$) and $\sigma_y=i(|e\rangle\langle g|-|g\rangle\langle e|)$ the Pauli matrix around $y$. Since $\langle a_\mathrm{in}\rangle,\langle b_\mathrm{in}\rangle\in\mathbb{R}$, the Rabi frequencies $\Omega_{a,b}$ are real numbers whose sign encodes the phase ($0$ or $\pi$) of the corresponding drive. This Hamiltonian represents only the qubit energy in presence of the drive but does not capture the individual energy exchanges between the qubit and the propagating modes. To access it, one needs to express the outgoing photon flux in each line, which corresponds to the outgoing power normalized by the qubit photon energy $\hbar\omega_q$. Using Eq.~(\ref{equ:input}) it reads on average~\cite{Cottet2019} 
\begin{align}
\langle a_\mathrm{out}^\dagger a_\mathrm{out}\rangle & = \langle a_\mathrm{in}^\dagger a_\mathrm{in} \rangle + \gamma_a \frac{1+\langle \sigma_z\rangle}{2} - \frac{\Omega_a}{2} \langle \sigma_x \rangle
\label{equ:photons}\\
\langle b_\mathrm{out}^\dagger b_\mathrm{out}\rangle & = \langle b_\mathrm{in}^\dagger b_\mathrm{in} \rangle + \gamma_b \frac{1+\langle \sigma_z\rangle}{2} - \frac{\Omega_b}{2} \langle \sigma_x \rangle\,.
\end{align}
Note that $\langle \sigma_x\rangle$ appears in the above equations because we have chosen to restrict our analysis to the case where both drives in $a$ and $b$ are captured by a term in $\sigma_y$ in the Hamiltonian. The three terms of the outgoing power can be interpreted as follows. The first one is the outgoing input power reflected on the cavity. The second, proportional to the coupling rate of the qubit to the line and to the probability to find the qubit in the excited state corresponds to spontaneous emission in the line. The last term results from the interference between the input drive (through $\Omega_{a,b}$) and the fluorescence field emitted by the qubit. It represents stimulated emission or absorption. It is this interference effect that gives rise to directionality. For two cancelling drives let us take the convention $\Omega_a = -\Omega_b = \Omega > 0$. The qubit Hamiltonian simplifies as $H=0$ and the qubit only evolves according to decoherence, while the sign of $\langle\sigma_x\rangle$ controls the direction of the power transfer. If $\langle\sigma_x\rangle>0$, energy is taken from $a$ and transferred to $b$, and vice versa. In the absence of decoherence $\langle\sigma_x\rangle$ stays at the value $\sin\theta\cos\phi$ set by the prepared superposition $|\psi\rangle=\cos(\theta/2)|g\rangle+e^{i\phi}\sin(\theta/2)|e\rangle$. The stimulated contribution $\tfrac{\Omega}{2}\langle\sigma_x\rangle$ to the transferred power is therefore proportional to the qubit coherence: it reverses with the sign of $\langle\sigma_x\rangle$ and its magnitude is maximal for a maximally coherent state, $|\langle\sigma_x\rangle|=1$. This is the same coherence that sets the slope of the Rabi oscillation exploited in Ref.~\cite{Cottet2017} and the power delivered by measurement-driven engines~\cite{ElouardJordan2018,Elouard2017PRL}, and it identifies quantum coherence as the resource for the directional power transfer~\cite{Korzekwa2016,Monsel2020,Maffei2021}.

In practice, the output field of line $a$ is routed through a chain of amplifiers, starting with a near quantum-limited Josephson parametric amplifier~\cite{PhysRevX.6.011002}, and demodulated at room temperature into its two quadratures $I(t)$ and $Q(t)$. Averaging over many identical realizations, the measured power relates to the outgoing photon flux through
\begin{equation}
    \overline{I^2(t)+Q^2(t)} = G\,\langle a_\mathrm{out}^\dagger a_\mathrm{out}\rangle + \dot{n}_\mathrm{th}\,,
\end{equation}
where $G$ is the overall power gain of the amplification chain and $\dot{n}_\mathrm{th}$ is the rate of noise photons added by the detection, including the amplifier noise and the residual thermal occupation. The gain $G$ enters as a global multiplicative factor and the noise contribution $\dot{n}_\mathrm{th}$ as a constant offset; neither affects the determination of the rates or the phase dependence studied below. The sample, its parameters, and the calibration of the drives are described in chapter 3 in Ref.~\cite{Cottet2019}.

\section{\label{sec:purcell}Purcell rate measurement by fluorescence}
In a physical implementation, the qubit is not only coupled to the transmission lines but is also prone to decoherence due to additional loss channels such as dielectric loss. The scattering of propagating microwaves by such an artificial atom was demonstrated in Ref.~\cite{AstafievRF2010}. The coupling rate to a line is therefore always smaller than the total energy relaxation rate $\gamma_a<\gamma_1=(T_1)^{-1}$, where $T_1$ is the qubit lifetime. We similarly denote by $\gamma_2=(T_2)^{-1}$ the qubit decoherence rate, with $T_2$ the coherence time. In a fluorescence measurement experiment the ratio $\gamma_a/\gamma_1$ represents the maximal fraction of information that can be recovered about the state of the qubit by the measurement apparatus as it decays~\cite{Murch2013,PhysRevX.6.011002,Campagne-Ibarcq2016a}, therefore knowing this ratio is critical to optimizing the design and successfully reconstructing quantum trajectories. The measurement of the photon flux emitted by the qubit in the line allows us to finely access this quantity. Indeed, by sweeping the amplitude of the driving field $a_\mathrm{in}$, we can go from a regime where the photon flux is given by spontaneous emission and proportional to $\gamma_a$ (second term in Eq.~\ref{equ:photons}) to one dominated by stimulated emission and proportional to $\Omega_a$. The comparison between the two gives $\gamma_a$. A related interplay between the coherent and incoherent components of the field scattered by a driven emitter has been exploited to extract the decay and decoherence rates of a superconducting qubit~\cite{Lu2021} and to perform primary thermometry of propagating microwave modes~\cite{Scigliuzzo2020}.

\begin{figure}[h!]
\centering
\includegraphics[width=0.8\linewidth]{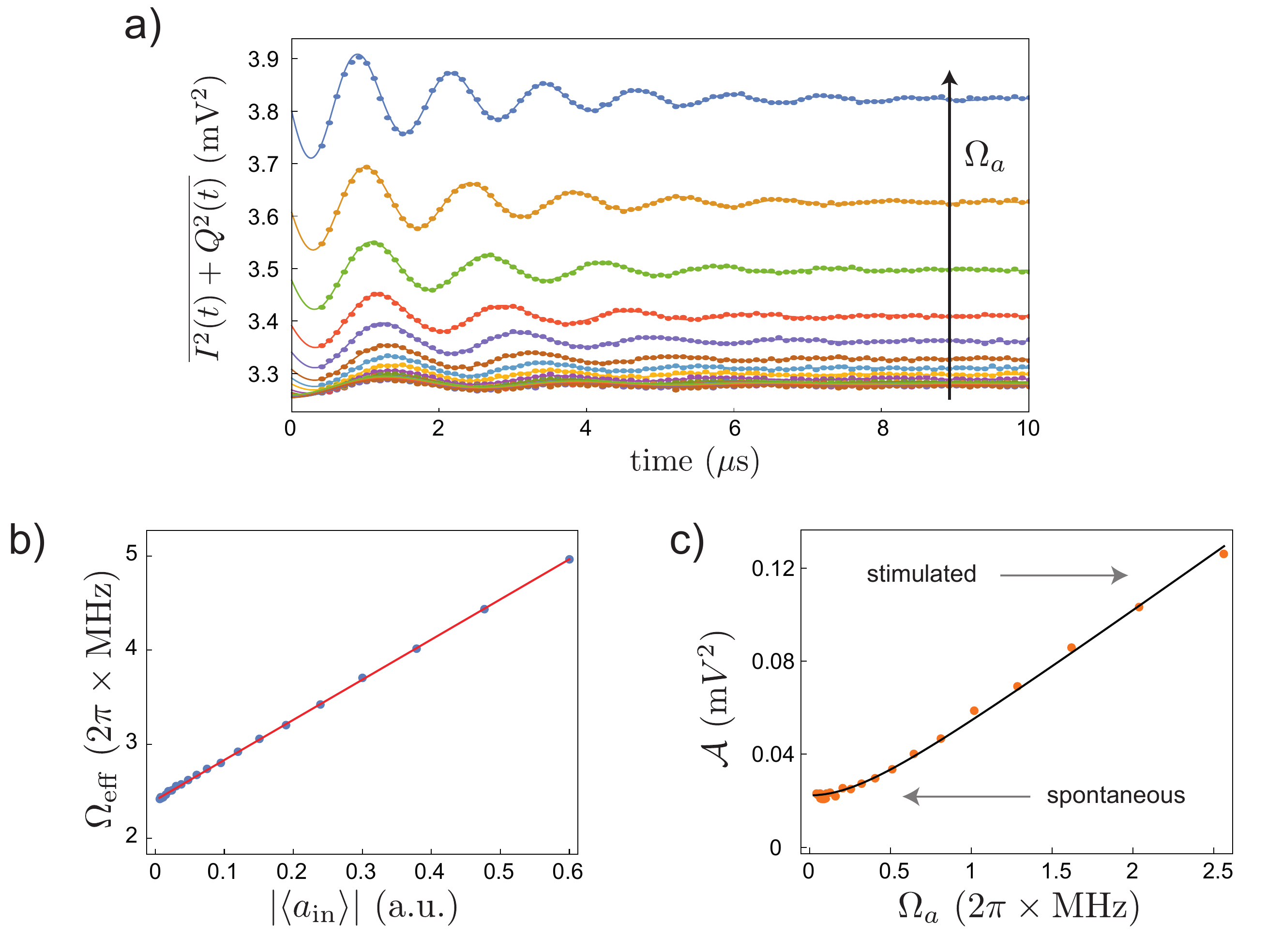}
\caption{\textbf{Purcell rate determination using power measurement of fluorescence.} (a) Time-resolved power measurement of fluorescence of the qubit under two Rabi drives in phase while sweeping the amplitude of the drive $\Omega_a$. The experimental data (points) are fitted by exponentially decaying sine functions (lines). (b) Evolution of the oscillation frequency $\Omega_\mathrm{eff}=\Omega_a+\Omega_b$ as a function of the experimental drive amplitude on $a$. The fit (red line) gives the scaling factor between the amplitude and $\Omega_a$. (c) The evolution of the amplitude of the oscillations as a function of $\Omega_a$ shows the transition from the spontaneous emission regime to the stimulated emission regime. The data (points) are fitted (black line) to find the Purcell rate $\gamma_a=72$~kHz.}
\label{Fig2}
\end{figure}

To measure $\gamma_a$, we drive the qubit with two \emph{in-phase} Rabi drives in lines $a$ and $b$. In contrast with the cancelling configuration used in the rest of the paper, the two drives now add up, leading to the qubit Hamiltonian $H=-\hbar\Omega_\mathrm{eff}\sigma_y/2$ with $\Omega_\mathrm{eff}=\Omega_a+\Omega_b$. We sweep $\Omega_a$ from $0$ to $\Omega_a\gg\gamma_1$ while keeping $\Omega_b$ constant and large enough that $\Omega_b\gg\gamma_1,\gamma_2$. This ensures that the qubit undergoes damped Rabi oscillations at the frequency $\Omega_\mathrm{eff}$ and damping rate $(\gamma_1+\gamma_2)/2$. The time-resolved measured power is represented in Fig.~\ref{Fig2}a for various driving amplitudes $\Omega_a$. Each trace is fitted by a sine wave decaying at a rate $(\gamma_1+\gamma_2)/2$ with the phase, amplitude, frequency and offset of the oscillations as fit parameters (plain lines in Fig.~\ref{Fig2}a). The oscillation frequency $\Omega_\mathrm{eff}$ scales linearly with the drive amplitude $|\langle a_\mathrm{in}\rangle|$ and gives the scaling factor between the uncalibrated drive amplitude set in the experiment and the physical Rabi rate $\Omega_a$ (see Fig.~\ref{Fig2}b) as well as $\Omega_b=2\pi\times2.4$~MHz. 

The amplitude of the oscillations is captured by Eq.~(\ref{equ:photons}), and results from two distinct effects. Spontaneous emission is proportional to $\langle \sigma_z\rangle(t)$ and stimulated emission to $\langle \sigma_x\rangle(t)$. Since they evolve in phase quadrature, the amplitude is given by $\mathcal{A}=\mathcal{C}\sqrt{\gamma_a^2+\Omega_a^2}$ with $\mathcal{C}$ a contrast factor coming from the thermal excitation of the qubit and the global gain of the amplification chain. The measured amplitudes (Fig.~\ref{Fig2}c) show the transition from a photon flux dominated by spontaneous emission when $\Omega_a\ll\gamma_a$ with constant amplitude, to one dominated by stimulated emission when $\Omega_a\gg\gamma_a$ with amplitude scaling linearly with $\Omega_a$. The fit of the amplitudes yields $\gamma_a=2\pi\times72$~kHz. Note that this method for determining $\gamma_a$ does not require knowing the gain of the amplification chain, which simply acts as a scaling factor.

\section{\label{sec:routing}Power routing by quantum coherence}

\begin{figure}[h!]
\centering
\includegraphics[width=0.8\textwidth]{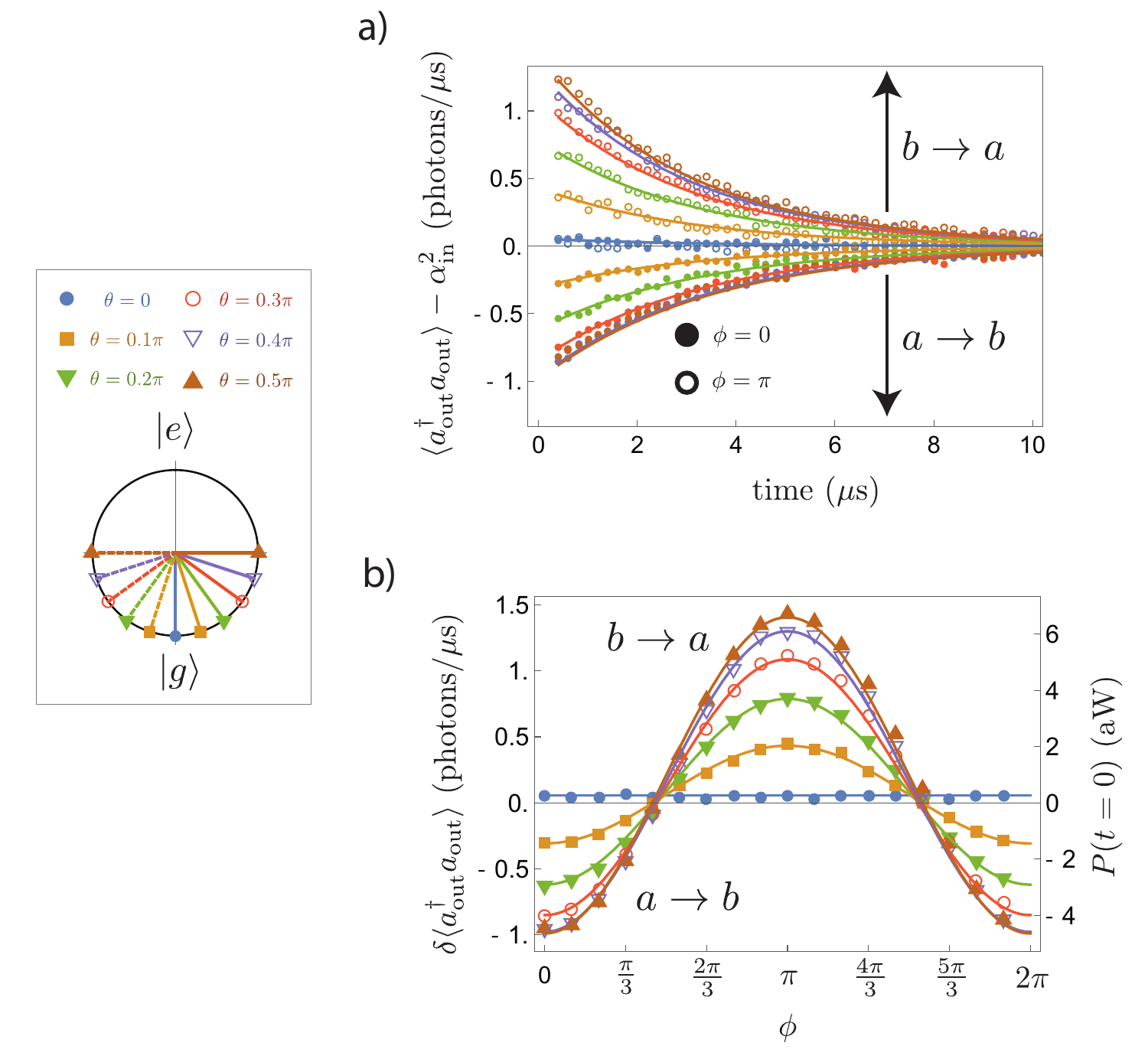}
\caption{\textbf{Evidence of power transfer.} (left panel) The qubit is initialized in various superpositions of states with phases $\phi=0$ (plain line) and $\phi=\pi$ (dashed line) represented in a 2D projection of the Bloch sphere, before turning on the two drives. a) Time-resolved measurement of the difference between the output and input photon flux in line $a$, showing directionality depending on the phase (full circles for $\phi=0$ and empty circles for $\phi=\pi$). The experimental data (points) is fitted with the theory (plain line) using the gain and a delay between the drives $\tau_d=18$~ns as fitting parameters. b) Control of the transfer direction by the phase of the superposition of states. The amplitude of the power transfer at $t=0$ (points) evolves as $\cos(\phi)$ and is well reproduced by the theory (plain lines).}
\label{Fig3}
\end{figure}
We now turn to directional transfer of power from one drive to the other, and start by demonstrating that directionality is indeed controlled by the phase of the qubit superposition of states, that is by its quantum coherence. In all the following, the drives are canceling each other out, meaning that the qubit dynamics is only due to decoherence. The qubit is initialized in a superposition of states before turning on the drives with $\Omega_a=-\Omega_b=2\pi\times0.41$~MHz. Ideally we would prepare $|\psi\rangle=\cos(\theta/2)|g\rangle+e^{i\phi}\sin(\theta/2)|e\rangle$, but the fidelity of preparation is reduced by the initial thermal state of the qubit, which has a finite excited state probability $P_e^\mathrm{th}=3.6\%$. The difference between the outgoing photon flux $\langle a_\mathrm{out}^\dagger a_\mathrm{out}\rangle$ and the input drive $\langle a_\mathrm{in}^\dagger a_\mathrm{in}\rangle=\alpha_\mathrm{in}^2$ is represented in Fig.~\ref{Fig3}a for various values of $\theta$ and $\phi=0$ or $\pi$ (see left panel). It exhibits a clear signature of directionality: when $\phi=0$, less energy is collected in line $a$ than initially sent, $\langle a_\mathrm{out}^\dagger a_\mathrm{out}\rangle\leq\langle a_\mathrm{in}^\dagger a_\mathrm{in}\rangle$, meaning that part of it has been transferred to $b$; when $\phi=\pi$, more energy is collected, meaning that energy has been transferred from $b$ to $a$. Importantly, the curves at $\phi=0$ and $\phi=\pi$ are not symmetrical because of spontaneous emission, which is not directional, hence always emits no matter the phase of the superposition. The amount of power emitted by the qubit in the line, hence the power transfer, decays in time as the qubit decoheres. Since the two cancelling drives leave $H=0$, the qubit undergoes pure relaxation and
\begin{align}
\langle\sigma_z\rangle(t) &= \langle\sigma_z\rangle_\mathrm{th} + \left(\langle\sigma_z\rangle_0 - \langle\sigma_z\rangle_\mathrm{th}\right)e^{-\gamma_1 t}\,,\label{equ:sz}\\
\langle\sigma_x\rangle(t) &= \langle\sigma_x\rangle_0\,e^{-\gamma_2 t}\,,\label{equ:sx}
\end{align}
with $\langle\sigma_x\rangle_0=\sin\theta\cos\phi$ set by the prepared superposition and $\langle\sigma_z\rangle_\mathrm{th}=2P_e^\mathrm{th}-1$ the thermal steady value. Injected into Eq.~(\ref{equ:photons}), these expressions show that the spontaneous emission term, proportional to $\langle\sigma_z\rangle$, decays at the rate $\gamma_1$ ($T_1=2.22\ \mu$s) while the stimulated emission term, proportional to $\langle\sigma_x\rangle$, decays at the rate $\gamma_2$ ($T_2=3.21\ \mu$s). The qubit eventually reaches thermal equilibrium and the outgoing power equals the one that was sent. The data points are in good agreement with the theory of Eq.~(\ref{equ:photons}) with only two fit parameters: the global power gain of the amplification chain and the presence of a small delay $\tau_d=18$~ns between the times at which the two drives reach the qubit. The qubit therefore undergoes a small rotation in the Bloch sphere before emitting in the lines, which explains why it emits a small amount of power even when initialized in thermal equilibrium (blue curve).

The amount of power transferred should depend continuously on the phase of the superposition $\phi$, as it is sensitive to the expectation value $\langle\sigma_x\rangle\propto\sin(\theta)\cos(\phi)$. To test it, we measure the outgoing power for different values of $\phi$ and $\theta$, extract the amplitude of the power transfer $\delta\langle a_\mathrm{out}^\dagger a_\mathrm{out}\rangle=\langle a_\mathrm{out}^\dagger a_\mathrm{out}\rangle(t=0)-\langle a_\mathrm{in}^\dagger a_\mathrm{in}\rangle$, and plot it as a function of $\phi$ (Fig.~\ref{Fig3}b). As expected it evolves as $\cos(\phi)$ and the system exhibits a continuous transition from a power transfer from $a$ to $b$ to one from $b$ to $a$. The amplitude increases with $\sin(\theta)$, that is with the qubit coherence, and is maximal for the maximally coherent state $\theta=\pi/2$. As the qubit is prepared in states containing more excited-state population, spontaneous emission increases and the curves are shifted up. This upward shift does not mean that the transfer is more efficient from $b$ to $a$ than from $a$ to $b$: it simply reflects the spontaneous emission of the qubit, which adds the same phase-independent contribution to line $a$ for $\phi=0$ and $\phi=\pi$. The curves are well reproduced by the theory with no fit parameters. The measured photon flux is converted into power using the photon energy $\hbar\omega_q$ and corresponds to a differential power at $t=0$ between $-4$ and $+6$~aW (Fig.~\ref{Fig3}b, right axis).

\section{\label{sec:efficiency}Photon routing efficiency}
\begin{figure}[h!]
\centering
\includegraphics[width=0.9\linewidth]{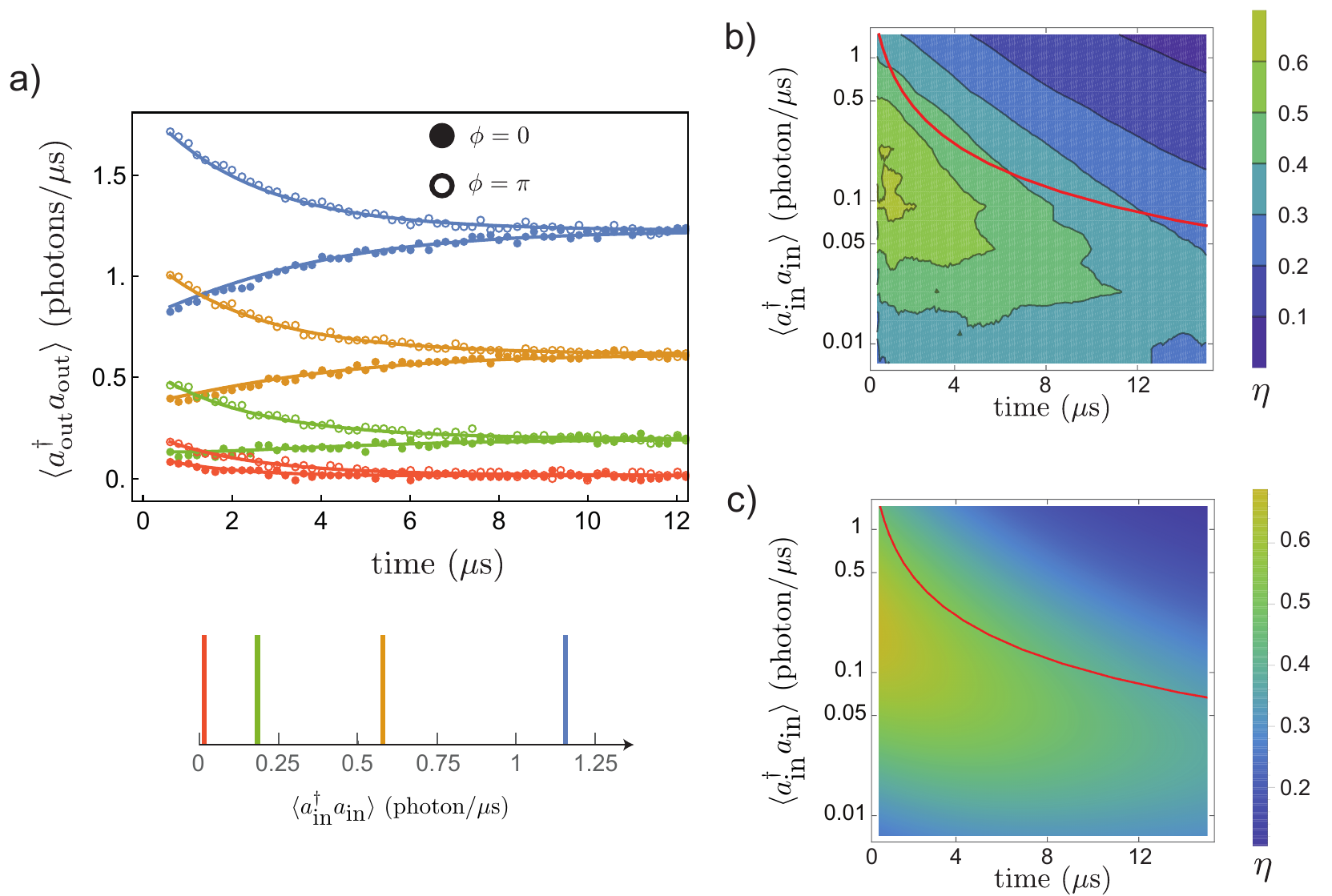}
\caption{\textbf{Photon transfer efficiency.} (a) Evolution of the outgoing photon flux measured in $a$ as a function of time for a qubit initialized in $|g\rangle+e^{i\phi}|e\rangle$ with $\phi=0$ (full circles) or $\phi=\pi$ (empty circles) with various input photon fluxes (bottom panel). It displays the transition from a flux dominated by spontaneous emission (red) with weak directionality to one where the input drive overcomes the emission of the qubit (blue). The highest directionality is obtained in the intermediate regime (green). (b) Experimental photon transfer efficiency obtained by integrating the photon flux for a time $\tau$ and (c) ideal theoretical efficiency based on the qubit coherence times. The input fluxes and pulse times corresponding to 1 photon on average are represented by the red line. The maximum efficiency is measured at 63\% for $6\times10^{-2}$ input photons (ideal 69\%) and at 46\% for 1 input photon (ideal 54\%).}
\label{Fig4}
\end{figure}

If stimulated emission can successfully transfer power from one line to the other, it does so in competition with the reflected input drive and spontaneous emission. The total transfer therefore results from a trade-off between these three terms. When the input drive is weak, $\langle a^\dagger_\mathrm{in}a_\mathrm{in}\rangle\ll\gamma_a$, directionality is reduced because the photon flux is dominated by spontaneous emission. On the other hand, a strong drive such that $\langle a^\dagger_\mathrm{in}a_\mathrm{in}\rangle\gg\gamma_a$ overcomes both the spontaneous and the stimulated emission of the qubit, so that the reflected input drive dominates and again limits directionality. We illustrate this effect in Fig.~\ref{Fig4}a, where we initialize the qubit in $|g\rangle+e^{i\phi}|e\rangle,\ \phi=0,\pi$ and measure the photon flux in $a$ for various input photon powers (see inset). As expected, larger drive amplitudes increase the contribution of stimulated emission but it is made at the expense of a bigger offset. In order to reproduce theoretically (lines) the experimental points (circles), we have to assume a residual Rabi drive on the qubit whose amplitude is $3\%$ of the initial drive. We interpret this effect as a small drift between the calibration of the drives and the time at which the data were taken (see chapter 3 in Ref.~\cite{Cottet2019}).

In order to quantify the photon transfer efficiency one has to account for the total number of quanta in the system made of the two drives and the qubit. A perfect transfer would correspond to all the energy exiting in one of the two lines and perfect extinction in the other. In the case of a strongly asymmetric coupling to the lines the efficiency can be defined as the on/off ratio
\begin{equation}
    \eta=\frac{N_\pi-N_0}{N_\pi+N_0}
\end{equation}
where $N_\phi=\int_0^\tau\langle a_\mathrm{out}^\dagger a_\mathrm{out}\rangle_\phi\mathrm{d}t$ is the average number of photons collected in line $a$ after a time $\tau$ when the qubit is initialized with a phase $\phi$. Because of decoherence we expect the efficiency at a given drive amplitude to decrease monotonically in time. The experimental efficiency is represented in Fig.~\ref{Fig4}b as a function of time and the input photon flux. It is maximal at the smallest accessible time $\tau_m=0.61\ \mu$s and reaches a maximum of 63\% for $\langle a_\mathrm{in}^\dagger a_\mathrm{in}\rangle=0.092$~photons per $\mu$s. The ideal theoretical efficiency (Fig.~\ref{Fig4}c) based on the qubit $T_1$ and $T_2$ is close to the measured one and predicts a maximum of 69\%. The difference between the two maxima probably comes from the imperfect cancelation between the two drives. Reaching the maximum bound corresponds to sending a drive containing, on average, $\langle a_\mathrm{in}^\dagger a_\mathrm{in}\rangle\times\tau_m=6\times10^{-2}$ photons, which makes the success probability small. A more practical measure is the photon transfer efficiency when one input photon (on average) is sent on the qubit, represented by the red line in Fig.~\ref{Fig4}b and c. The experimental (resp. ideal) transfer efficiency reaches a maximum of 46\% (resp. 54\%) for a duration $\tau=2.7\ \mu$s.

\section{Conclusion}
\label{sec:conclusion}
We have shown that a single qubit driven simultaneously by two spatially separated modes can transfer power from one mode to the other, with a direction set by the phase $\phi$ of the qubit superposition rather than by any classical control parameter. The transferred power follows the qubit coherence $\langle\sigma_x\rangle$, reverses sign between $\phi=0$ and $\phi=\pi$, and competes with the non-directional contributions of the reflected drive and of spontaneous emission. Balancing these contributions yields a routing efficiency of $63\%$ for an equal superposition, close to the bound of $69\%$ fixed by the measured coherence times. The fluorescence power measurement used to extract this bound also provides a direct determination of the Purcell rate $\gamma_a$ into the detected line.
The efficiency reflects a trade-off between the energy that a qubit superposition can pump between two modes and the information that leaks to the environment, a connection that relates this power-routing scheme to the thermodynamics of measurement and to the energetics of work exchange explored with the same platform~\cite{Cottet2017,ElouardJordan2018,Elouard2017PRL,Monsel2020,Stevens2022,Dassonneville2026}.

It is instructive to ask what would change if the qubit were replaced by a doubly driven linear system. Consider a harmonic oscillator prepared in a coherent state $\alpha$ and driven on resonance by the two modes with opposite amplitudes, so that their net effect on the oscillator cancels, as in our experiment. A directional transfer of power between $a$ and $b$ is then observed as well, and with a suitable choice of phase reference its amplitude is proportional to $\mathrm{Re}\,\alpha$. The effect is in this case entirely classical, since a coherent state is the most classical state of the oscillator~\cite{Glauber1963,Sudarshan1963}. The comparison highlights a duality between the qubit and the oscillator. The classical states of the qubit, $|g\rangle$ and $|e\rangle$, map onto the quantum states of the oscillator, $|0\rangle$ and $|1\rangle$, while the superpositions of the qubit, $(|g\rangle\pm|e\rangle)/\sqrt{2}$, map onto the classical coherent states $\alpha$. What appears as a classical interference effect for a linear system thus relies, for a two-level system, on a coherent superposition of energy eigenstates with no classical counterpart.

\section*{Acknowledgements}
The authors warmly acknowledge P. Bertet, M. Brune, D. Carpentier, J.-M. Raimond, and C. Sayrin for enlightening discussions.


\paragraph{Funding information}
Nanofabrication has been made within the consortium Salle Blanche Paris Centre. This work was funded by Agence Nationale de la Recherche under the grant ANR-17-ERC2-0001-01.

\paragraph{Data availability}
The data that support the findings of this study are available from the corresponding author upon reasonable request.


\begin{thebibliography}{10}
\providecommand{\url}[1]{\texttt{#1}}
\providecommand{\urlprefix}{URL }
\expandafter\ifx\csname urlstyle\endcsname\relax
  \providecommand{\doi}[1]{doi:\discretionary{}{}{}#1}\else
  \providecommand{\doi}{doi:\discretionary{}{}{}\begingroup
  \urlstyle{rm}\Url}\fi
\providecommand{\eprint}[2][]{\url{#2}}

\bibitem{Binder2018a}
F.~Binder, L.~Correa, C.~Gogolin, J.~Anders and G.~Adesso, eds.,
\newblock \emph{{Thermodynamics in the Quantum Regime: Fundamental Aspects and
  New Directions}},
\newblock Springer (2018).

\bibitem{AuffevesElouard2026QuantumEnergetics}
A.~Auff\`eves and C.~Elouard,
\newblock \emph{Quantum energetics, foundations, applications},
\newblock In S.~Campbell \emph{et~al.}, eds., \emph{Roadmap on Quantum
  Thermodynamics}, vol.~11, p. 012501. IOP Publishing,
\newblock \doi{10.1088/2058-9565/ae1e27} (2026).

\bibitem{Campbell2026}
S.~Campbell \emph{et~al.},
\newblock \emph{{Roadmap on Quantum Thermodynamics}},
\newblock Quantum Science and Technology \textbf{11}(1), 012501 (2026),
\newblock \doi{10.1088/2058-9565/ae1e27}.

\bibitem{Timossi2018}
G.~F. Timossi, A.~Fornieri, F.~Paolucci, C.~Puglia and F.~Giazotto,
\newblock \emph{{Phase-Tunable Josephson Thermal Router}},
\newblock Nano Letters \textbf{18}(3), 1764 (2018),
\newblock \doi{10.1021/acs.nanolett.7b04906}.

\bibitem{Ronzani2018}
A.~Ronzani, B.~Karimi, J.~Senior, Y.-C. Chang, J.~T. Peltonen, C.~Chen and
  J.~P. Pekola,
\newblock \emph{{Tunable photonic heat transport in a quantum heat valve}},
\newblock Nature Physics \textbf{14}(10), 991 (2018),
\newblock \doi{10.1038/s41567-018-0199-4}.

\bibitem{Hoi2011}
I.-C. Hoi, C.~M. Wilson, G.~Johansson, T.~Palomaki, B.~Peropadre and
  P.~Delsing,
\newblock \emph{{Demonstration of a Single-Photon Router in the Microwave
  Regime}},
\newblock Physical Review Letters \textbf{107}(7), 073601 (2011),
\newblock \doi{10.1103/PhysRevLett.107.073601}.

\bibitem{Lodahl2017}
P.~Lodahl, S.~Mahmoodian, S.~Stobbe, A.~Rauschenbeutel, P.~Schneeweiss,
  J.~Volz, H.~Pichler and P.~Zoller,
\newblock \emph{{Chiral quantum optics}},
\newblock Nature \textbf{541}(7638), 473 (2017),
\newblock \doi{10.1038/nature21037}.

\bibitem{Aoki2009}
T.~Aoki, A.~S. Parkins, D.~J. Alton, C.~A. Regal, B.~Dayan, E.~Ostby, K.~J.
  Vahala and H.~J. Kimble,
\newblock \emph{{Efficient Routing of Single Photons by One Atom and a
  Microtoroidal Cavity}},
\newblock Physical Review Letters \textbf{102}(8), 083601 (2009),
\newblock \doi{10.1103/PhysRevLett.102.083601}.

\bibitem{Shomroni2014}
I.~Shomroni, S.~Rosenblum, Y.~Lovsky, O.~Bechler, G.~Guendelman and B.~Dayan,
\newblock \emph{{All-optical routing of single photons by a one-atom switch
  controlled by a single photon}},
\newblock Science \textbf{345}(6199), 903 (2014),
\newblock \doi{10.1126/science.1254699}.

\bibitem{Mitsch2014}
R.~Mitsch, C.~Sayrin, B.~Albrecht, P.~Schneeweiss and A.~Rauschenbeutel,
\newblock \emph{{Quantum state-controlled directional spontaneous emission of
  photons into a nanophotonic waveguide}},
\newblock Nature Communications \textbf{5}, 5713 (2014),
\newblock \doi{10.1038/ncomms6713}.

\bibitem{Sayrin2015}
C.~Sayrin, C.~Junge, R.~Mitsch, B.~Albrecht, D.~O'Shea, P.~Schneeweiss, J.~Volz
  and A.~Rauschenbeutel,
\newblock \emph{{Nanophotonic Optical Isolator Controlled by the Internal State
  of Cold Atoms}},
\newblock Physical Review X \textbf{5}(4), 041036 (2015),
\newblock \doi{10.1103/PhysRevX.5.041036}.

\bibitem{Scheucher2016}
M.~Scheucher, A.~Hilico, E.~Will, J.~Volz and A.~Rauschenbeutel,
\newblock \emph{{Quantum optical circulator controlled by a single chirally
  coupled atom}},
\newblock Science \textbf{354}(6319), 1577 (2016),
\newblock \doi{10.1126/science.aaj2118}.

\bibitem{Gheeraert2020}
N.~Gheeraert, S.~Kono and Y.~Nakamura,
\newblock \emph{{Programmable directional emitter and receiver of itinerant
  microwave photons in a waveguide}},
\newblock Physical Review A \textbf{102}(5), 053720 (2020),
\newblock \doi{10.1103/PhysRevA.102.053720}.

\bibitem{Guimond2020}
P.-O. Guimond, B.~Vermersch, M.~L. Juan, A.~Sharafiev, G.~Kirchmair and
  P.~Zoller,
\newblock \emph{{A unidirectional on-chip photonic interface for
  superconducting circuits}},
\newblock npj Quantum Information \textbf{6}(1), 32 (2020),
\newblock \doi{10.1038/s41534-020-0261-9}.

\bibitem{Redchenko2023}
E.~S. Redchenko, A.~V. Poshakinskiy, R.~Sett, M.~{\v{Z}}emli{\v{c}}ka, A.~N.
  Poddubny and J.~M. Fink,
\newblock \emph{{Tunable directional photon scattering from a pair of
  superconducting qubits}},
\newblock Nature Communications \textbf{14}(1), 2998 (2023),
\newblock \doi{10.1038/s41467-023-38761-6}.

\bibitem{Kannan2023}
B.~Kannan, A.~Almanakly, Y.~Sung, A.~Di~Paolo, D.~A. Rower,
  J.~Braum{\"{u}}ller, A.~Melville, B.~M. Niedzielski, A.~Karamlou, K.~Serniak,
  A.~Veps{\"{a}}l{\"{a}}inen, M.~E. Schwartz \emph{et~al.},
\newblock \emph{{On-demand directional microwave photon emission using
  waveguide quantum electrodynamics}},
\newblock Nature Physics \textbf{19}(3), 394 (2023),
\newblock \doi{10.1038/s41567-022-01869-5}.

\bibitem{Luneau2022}
J.~Luneau, C.~Dutreix, Q.~Ficheux, P.~Delplace, B.~Dou{\c{c}}ot, B.~Huard and
  D.~Carpentier,
\newblock \emph{{Topological power pumping in quantum circuits}},
\newblock Physical Review Research \textbf{4}(1), 013169 (2022),
\newblock \doi{10.1103/PhysRevResearch.4.013169},
\newblock \eprint{2109.12897}.

\bibitem{Aamir2025}
M.~A. Aamir, P.~Jamet~Suria, J.~A. Mar{\'{i}}n~Guzm{\'{a}}n,
  C.~Castillo-Moreno, J.~M. Epstein, N.~Yunger~Halpern and S.~Gasparinetti,
\newblock \emph{{Thermally driven quantum refrigerator autonomously resets a
  superconducting qubit}},
\newblock Nature Physics \textbf{21}(2), 318 (2025),
\newblock \doi{10.1038/s41567-024-02708-5},
\newblock \eprint{2305.16710}.

\bibitem{Maffei2021}
M.~Maffei, P.~A. Camati and A.~Auff{\`{e}}ves,
\newblock \emph{{Probing nonclassical light fields with energetic witnesses in
  waveguide quantum electrodynamics}},
\newblock Physical Review Research \textbf{3}(3), L032073 (2021),
\newblock \doi{10.1103/PhysRevResearch.3.L032073}.

\bibitem{Wenniger2023}
I.~Maillette~de Buy~Wenniger, S.~E. Thomas, M.~Maffei, S.~C. Wein, M.~Pont,
  N.~Belabas, S.~Prasad, A.~Harouri, A.~Lema{\^{i}}tre, I.~Sagnes, N.~Somaschi,
  A.~Auff{\`{e}}ves \emph{et~al.},
\newblock \emph{{Experimental Analysis of Energy Transfers between a Quantum
  Emitter and Light Fields}},
\newblock Physical Review Letters \textbf{131}(26), 260401 (2023),
\newblock \doi{10.1103/PhysRevLett.131.260401}.

\bibitem{Prasad2024}
S.~P. Prasad, M.~Maffei, P.~A. Camati, C.~Elouard and A.~Auff{\`{e}}ves,
\newblock \emph{{Thermodynamics of autonomous optical Bloch equations}},
\newblock SciPost Physics \textbf{20}(4), 112 (2026),
\newblock \doi{10.21468/SciPostPhys.20.4.112},
\newblock \eprint{2404.09648}.

\bibitem{Monsel2020}
J.~Monsel, M.~Fellous-Asiani, B.~Huard and A.~Auff{\`{e}}ves,
\newblock \emph{{The Energetic Cost of Work Extraction}},
\newblock Physical Review Letters \textbf{124}(13), 130601 (2020),
\newblock \doi{10.1103/PhysRevLett.124.130601}.

\bibitem{ElouardJordan2018}
C.~Elouard and A.~N. Jordan,
\newblock \emph{{Efficient Quantum Measurement Engines}},
\newblock Physical Review Letters \textbf{120}(26), 260601 (2018),
\newblock \doi{10.1103/PhysRevLett.120.260601}.

\bibitem{Elouard2017PRL}
C.~Elouard, D.~Herrera-Mart{\'{i}}, B.~Huard and A.~Auff{\`{e}}ves,
\newblock \emph{{Extracting Work from Quantum Measurement in Maxwell's Demon
  Engines}},
\newblock Physical Review Letters \textbf{118}(26), 260603 (2017),
\newblock \doi{10.1103/PhysRevLett.118.260603}.

\bibitem{Dassonneville2026}
R.~Dassonneville, C.~Elouard, R.~Cazali, R.~Assouly, A.~Bienfait,
  A.~Auff{\`{e}}ves and B.~Huard,
\newblock \emph{{Amplifying microwave pulses with a single qubit engine fueled
  by quantum measurements}},
\newblock Physical Review Research \textbf{8}(1), 013228 (2026),
\newblock \doi{10.1103/rygc-bc3c},
\newblock \eprint{2501.17069}.

\bibitem{Cottet2017}
N.~Cottet, S.~Jezouin, L.~Bretheau, P.~Campagne-Ibarcq, Q.~Ficheux, J.~Anders,
  A.~Auff{\`{e}}ves, R.~Azouit, P.~Rouchon and B.~Huard,
\newblock \emph{{Observing a quantum Maxwell demon at work}},
\newblock Proceedings of the National Academy of Sciences \textbf{114}(29),
  7561 (2017),
\newblock \doi{10.1073/pnas.1704827114}.

\bibitem{Potts2025}
M.~Schrauwen, A.~Daniel, M.~Janovitch and P.~P. Potts,
\newblock \emph{{Thermodynamic Framework for Coherently Driven Systems}},
\newblock Physical Review Letters \textbf{135}(22), 220201 (2025),
\newblock \doi{10.1103/zdbv-rksc},
\newblock \eprint{2505.08558}.

\bibitem{Aziz2025}
F.~Aziz, K.-T. Lin, P.-Y. Wen, {Samina}, Y.-C. Lin, E.~Wiegand, C.-P. Lee,
  Y.-T. Cheng, Y.~Lu, C.-Y. Chen, C.-H. Chien, K.-M. Hsieh \emph{et~al.},
\newblock \emph{{Nearly quantum-limited microwave amplification via interfering
  degenerate stimulated emission in a single artificial atom}},
\newblock npj Quantum Information \textbf{11}(1), 45 (2025),
\newblock \doi{10.1038/s41534-025-00993-3}.

\bibitem{Paik2011a}
H.~Paik, D.~I. Schuster, L.~S. Bishop, G.~Kirchmair, G.~Catelani, a.~P. Sears,
  B.~R. Johnson, M.~J. Reagor, L.~Frunzio, L.~I. Glazman, S.~M. Girvin, M.~H.
  Devoret \emph{et~al.},
\newblock \emph{{Observation of High Coherence in Josephson Junction Qubits
  Measured in a Three-Dimensional Circuit QED Architecture}},
\newblock Physical Review Letters \textbf{107}(24), 240501 (2011),
\newblock \doi{10.1103/PhysRevLett.107.240501}.

\bibitem{Gardiner2004}
C.~Gardiner and P.~Zoller,
\newblock \emph{{Quantum Noise}},
\newblock Springer-Verlag Berlin Heidelberg, 3 edn.,
\newblock ISBN 978-3-540-22301-6 (2004).

\bibitem{Roy2018}
A.~Roy and M.~Devoret,
\newblock \emph{{Introduction to parametric amplification of quantum signals
  with Josephson circuits}},
\newblock Comptes Rendus Physique \textbf{17}(7), 740 (2016),
\newblock \doi{10.1016/j.crhy.2016.07.012}.

\bibitem{PhysRevX.6.011002}
P.~Campagne-Ibarcq, P.~Six, L.~Bretheau, A.~Sarlette, M.~Mirrahimi, P.~Rouchon
  and B.~Huard,
\newblock \emph{{Observing Quantum State Diffusion by Heterodyne Detection of
  Fluorescence}},
\newblock Phys. Rev. X \textbf{6}(1), 11002 (2016),
\newblock \doi{10.1103/PhysRevX.6.011002}.

\bibitem{Cottet2019}
N.~Cottet,
\newblock \emph{{\href{https://tel.archives-ouvertes.fr/tel-02002463}{Energy
  and Information in Fluorescence with Superconducting Circuits}}},
\newblock Ph.D. thesis, {\'E}cole Normale Sup{\'e}rieure, Paris (2019).

\bibitem{Korzekwa2016}
K.~Korzekwa, M.~Lostaglio, J.~Oppenheim and D.~Jennings,
\newblock \emph{{The extraction of work from quantum coherence}},
\newblock New Journal of Physics \textbf{18}(2), 023045 (2016),
\newblock \doi{10.1088/1367-2630/18/2/023045}.

\bibitem{AstafievRF2010}
O.~Astafiev, A.~M. Zagoskin, A.~A. Abdumalikov, Y.~A. Pashkin, T.~Yamamoto,
  K.~Inomata, Y.~Nakamura and J.~S. Tsai,
\newblock \emph{{Resonance Fluorescence of a Single Artificial Atom}},
\newblock Science \textbf{327}(5967), 840 (2010),
\newblock \doi{10.1126/science.1181918}.

\bibitem{Murch2013}
K.~W. Murch, S.~J. Weber, K.~M. Beck, E.~Ginossar and I.~Siddiqi,
\newblock \emph{{Reduction of the radiative decay of atomic coherence in
  squeezed vacuum}},
\newblock Nature \textbf{499}(7456), 62 (2013),
\newblock \doi{10.1038/nature12264}.

\bibitem{Campagne-Ibarcq2016a}
P.~Campagne-Ibarcq, S.~Jezouin, N.~Cottet, P.~Six, L.~Bretheau, F.~Mallet,
  A.~Sarlette, P.~Rouchon and B.~Huard,
\newblock \emph{{Using Spontaneous Emission of a Qubit as a Resource for
  Feedback Control}},
\newblock Physical Review Letters \textbf{117}(6), 060502 (2016),
\newblock \doi{10.1103/PhysRevLett.117.060502}.

\bibitem{Lu2021}
Y.~Lu, A.~Bengtsson, J.~J. Burnett, E.~Wiegand, B.~Suri, P.~Krantz, A.~F.
  Roudsari, A.~F. Kockum, S.~Gasparinetti, G.~Johansson and P.~Delsing,
\newblock \emph{{Characterizing decoherence rates of a superconducting qubit by
  direct microwave scattering}},
\newblock npj Quantum Information \textbf{7}(1), 35 (2021),
\newblock \doi{10.1038/s41534-021-00367-5}.

\bibitem{Scigliuzzo2020}
M.~Scigliuzzo, A.~Bengtsson, J.-C. Besse, A.~Wallraff, P.~Delsing and
  S.~Gasparinetti,
\newblock \emph{{Primary Thermometry of Propagating Microwaves in the Quantum
  Regime}},
\newblock Physical Review X \textbf{10}(4), 041054 (2020),
\newblock \doi{10.1103/PhysRevX.10.041054},
\newblock \eprint{2003.13522}.

\bibitem{Stevens2022}
J.~Stevens, D.~Szombati, M.~Maffei, C.~Elouard, R.~Assouly, N.~Cottet,
  R.~Dassonneville, Q.~Ficheux, S.~Zeppetzauer, A.~Bienfait, A.~N. Jordan,
  A.~Auffèves \emph{et~al.},
\newblock \emph{Energetics of a {Single} {Qubit} {Gate}},
\newblock Physical Review Letters \textbf{129}(11), 110601 (2022),
\newblock \doi{10.1103/PhysRevLett.129.110601},
\newblock ArXiv: 2109.09648.

\bibitem{Glauber1963}
R.~J. Glauber,
\newblock \emph{{Coherent and Incoherent States of the Radiation Field}},
\newblock Physical Review \textbf{131}(6), 2766 (1963),
\newblock \doi{10.1103/PhysRev.131.2766}.

\bibitem{Sudarshan1963}
E.~C.~G. Sudarshan,
\newblock \emph{{Equivalence of Semiclassical and Quantum Mechanical
  Descriptions of Statistical Light Beams}},
\newblock Physical Review Letters \textbf{10}(7), 277 (1963),
\newblock \doi{10.1103/PhysRevLett.10.277}.

\end{thebibliography}
\end{document}